# A Digital Game Maturity Model (DGMM)

***Saiqa Aleem, Luiz Fernando Capretz, Faheem Ahmed***

ABSTRACT

Game development is an interdisciplinary concept that embraces artistic, software engineering, management, and business disciplines. This research facilitates a better understanding of important dimensions of digital game development methodology. Game development is considered as one of the most complex tasks in software engineering. The increased popularity of digital games, the challenges faced by game development organizations in developing quality games, and high competition in the digital game industry demand a game development maturity assessment. Consequently, this study presents a Digital Game Maturity Model to evaluate the current development methodology in an organization. The framework of this model consists of assessment questionnaires, a performance scale, and a rating method. The main goal of the questionnaires is to collect information about current processes and practices. In general, this research contributes towards formulating a comprehensive and unified strategy for game development maturity evaluation. Two case studies were conducted and their assessment results reported. These demonstrate the level of maturity of current development practices in two organizations.

## 1. Introduction

The digital game market throughout the world has grown by over 7%–8% annually and has reached annual sales of around $5.5 billion in 2015 [1]. Newzoo [2] has also reported that the world-wide digital game market will reach $113.3 billion by 2018. The tremendous growth rate of the digital game market makes it obvious that game technology is easily accessible and has become more convenient. As a result, more and more people like to play games and have become motivated to design their own games. Furthermore, the game industry is so innovative that any technological advances either in hardware or software are applied to games before being adopted by other scientific fields [2],[3]. This remarkable growth of the digital game industry is capturing everyone's attention and also contributing to economic growth on a national level. Digital games are software applications that are installed on hardware devices such as video game consoles, computers, handheld devices, and PDAs. Digital game development involves multidisciplinary activities, which make it a complex task that is different from traditional software development. The multidisciplinary nature of the processes involved, which combine sound, art, control systems, artificial intelligence, and human factors, distinguish game development from other types of software development. It has become critically important to improve the game development process to address the issues faced by game development organizations in developing high-quality games to remain competitive and meet their financial objectives.

Digital game development is a complex task requiring real-time, high-quality performance. A number of game development tasks like real-time audio playback, high display frame rate, and fast processor response have an impact on game performance. Game programming is another difficult task that requires expert programmers dealing with thousands of lines of code. The variety of multi-disciplinary tasks, the low level of programming, and the large size of programs demand straightforward documentation, flexible design, and sustainable implementation. These will also help to ensure effective collaboration among the various development groups and to expedite future developments. Consequently, game developers need best practices guidelines and an assessment model to deal with the challenges they face in carrying out current processes. Ultimately, this will help them to improve current practices and will enable them to achieve high-quality levels.

Many researchers have discussed game development challenges. Pertillo *et al.* [5] surveyed the problems faced by game development organizations. The problems identified are categorized into four groups: scheduling problems, quality problems, budget problems, and management- and business-related problems. The overall game development process combines both an engineering process and the creation of artistic assets. Ramadan and Widyani [6] compared various game development strategies from a management perspective, and some researchers [7], [8], [9] have proposed frameworks for game development. To manage the development lifecycle effectively, guidelines based on best practices are required. Pertillo and Pimenta [10] highlighted the presence of agile practices in game development processes. Tschang [11] and Petrillo *et al.* [5] highlighted the issues in the game development process and its differences from traditional software development practices.

Management of development-team members and their interaction is critically important in this aspect. Hullett *et al.* [12], [13] have provided data analytics and empirical analysis of the game development process and discussed issues of interdisciplinary team involvement. Best practices in game development must consider certain elements such as staying on budget, timing, and producing the desired output. To assess game quality, five usability and quality criteria (functional, internally complete, balanced, fun, and accessible) can be used, but a process maturity model specific to the game development process is still needed to measure these processes for better management and high performance.

Profitable game development is another challenging endeavour. Every year, over 15,000 games are published and compete for players' attention and time [14]. This competitiveness of the global market and high cost of developing good-quality games are reasons for the digital game industry to improve its development processes. No study to date has directly addressed the assessment and improvement of the digital game development process to produce high-quality games, which would be ultimately beneficial for any country's economy. The process maturity models such as CMM [15] and CCMI [16] proposed by researchers in the past can be used to assess maturity level, but they also provide guidelines from a general perspective. However, game development is different from traditional software development and faces many specific challenges, as discussed above. There is also a need to capture other important perspectives of digital games such as the business, developers, and consumers. This effort will ultimately help to assess the maturity level of game development processes in an organization and provide guidelines to improve them. Researchers so far have focussed only on the technical aspects of game development, without considering improvements to the game development process that would ultimately be beneficial to digital game performance in the market.

This research introduces a first Digital Game Maturity Model (DGMM) for game development processes. Recently, game development organizations have been facing strong pressure to gain and retain competitive advantage. They need to identify different ways to control budgets, reduce time to market, and improve quality. In particular, a DGMM is made up of assessment questionnaires and a rating methodology. The assessment questionnaires contain factors that have been selected from literature review and three empirical investigations carried out by the authors based on three game-development perspectives: developer, consumer, and business [17]. The objective of the assessment questionnaires is to collect information about game development processes. As a proof of concept, proposed DGMM has been applied to two game development organizations, yielding results that are discussed in subsequent sections. The proposed framework will help organizations build the capability to identify gaps and bottlenecks in their current processes. This paper argues that the proposed DGMM can also help game development organizations to improve their business model because it captures the business perspective in addition to others. This DGMM can also help organizations to identify their current and target positions along with a roadmap to improve the current position towards the target position.

### 1.1 Research Motivation

SECOR [18] reported that the digital game industry, especially video games, will generate world-wide revenues of up to US $86.8 billion by 2016. It is clearly recommended in the report that Ontario game developers must demonstrate business acumen and management practices in development processes. The main challenge to developers is to achieve a balance between developing high-quality games and remaining within fixed budgets and deadlines. In the global digital game industry, Canada is a major player, both in terms of size and of quality of talent and resources. For these reasons, Canada has already replaced the United Kingdom in the third rank of computer game producers around the world [19]. To establish a leading game development industry (GDI) in Canada and attract foreign investment as proposed by the Ontario 2012 report [18], all stakeholders must strive to follow a process model to engineer games of the highest quality.

Lack of research in this area has also provided a motivation to select this domain. Hagan *et al.* [20] performed a systematic literature review of game development processes. They addressed three research questions:

- "What are the software process models in game development?"
- "What are the software process improvement (SPI) initiatives used in game development?"
- "What factors influence the adoption of software process models and SPI in game development?"

A systematic literature review demonstrated that most development companies used mixed, agile, or *ad-hoc* approaches and that no proper SPI initiatives were adopted. They also highlighted the importance of developing best-practice models for game development. Actually, most of the literature related to game development is found in the grey literature such as magazines, developer blogs, game development groups, and Web sites of game companies. This observation lends added importance to investigating what is actually happening in game development. Vargas *et al.* [21] performed a systematic mapping study on the quality of serious games. They investigated the particular quality characteristics of serious games and the methods used to investigate quality. The quality characteristics extracted from their research were effectiveness, efficiency, satisfaction, freedom from risk, and context coverage. They also identified product quality characteristics and revealed that no studies were found for compatibility, resource utilization, maintainability, and maturity assessment of games. They also suggested the urgent need for a quality assurance method that can evaluate game quality from the early stages of development and can be applied to any game.

The proposed digital game maturity model would capture various stakeholders' perspectives and would help game development organizations to improve the quality of their final product, i.e., the digital game. One of the main contributions of this research is to be the first digital game maturity model based on four dimensions that helps streamline current game development processes, and may also facilitate stakeholders to make correct decisions. This research is exceedingly beneficial to the digital game industry and also fills the research gap of concise game development process improvement guidelines and process assessment tools.

## 2. Literature Review

### *2.1 Game Development Process*

The digital game domain covers a great variety of player modes and genres [22], [23], [24]. The complexity of digital games has posed many challenges and issues in software development because it involves diverse activities in creative arts disciplines (storyboarding, design, refinement of animations, artificial intelligence, video production, scenarios, sound effects, marketing, and finally sales) besides technological and functional requirements [25]. This inherent diversity leads to a greatly fragmented domain from the perspectives of both underlying theory and design methodology. The digital game literature published in recent years has focussed mainly on technical issues. Issues of game production, development, and testing reflect only the general state of the art in software engineering. Pressman [26] stated that a game is a kind of software which entertains its users, but game development faces many challenges and issues if only a traditional software-development process is followed [27],[5].

Kanode and Haddad [27] stated that an important incorrect assumption has been made that game development follows the waterfall method. More recently, researchers have agreed that it must follow the incremental model because it combines the waterfall method with an iterative process. Petrillo et al. [5] reported a major concern, that developers for software creation in the game industry commonly use very poor development methodologies. The game development life cycle (GDLC) is the object of questions on many forms, which attempt to determine what types of practices are used. However, this question has no single answer. Few researchers have explored GDLC practices and then tried to answer questions like, "what are the phases of the GDLC?" Blitz Game Studios [28] proposed six phases for the GDLC, including pitch (initial design and game concept), pre-production (game design document), main production (implementation of game concepts), alpha (internal testing), beta (third-party testing), and finally the master phase (game launch). Hendrick [29] proposed a five-phase GDLC consisting of prototype (initial design prototype), pre-production (design document), production (asset creation, source code, integration aspects), beta (user feedback), and finally the live phase (ready to play). McGrath [30] divided the GDLC into seven phases: design (initial design and game design document), develop/redevelop (game engine development), evaluate (if not passed, then redevelop), test (internal testing), review release (third-party testing), and finally release (game launch).

Another GDLC proposed by Chandler [31] consisted of four phases: pre-production (design document and project planning), production (technical and artistic), testing (bug fixing), and finally post-production (post-release activities). The latest GDLC proposed in 2013 by Ramadan and Widyani [6] was based on the four GDLCs previously described. They proposed six phases, including initiation (rough concept), pre-production (creation of game design and prototype), production (formal details, refinement, implementation), testing (bug reports, refinement testing, change requests), beta (third-party testing), and finally, release (public release). Fig. 1 shows the general phases of the game development process and a related list of key process activities.

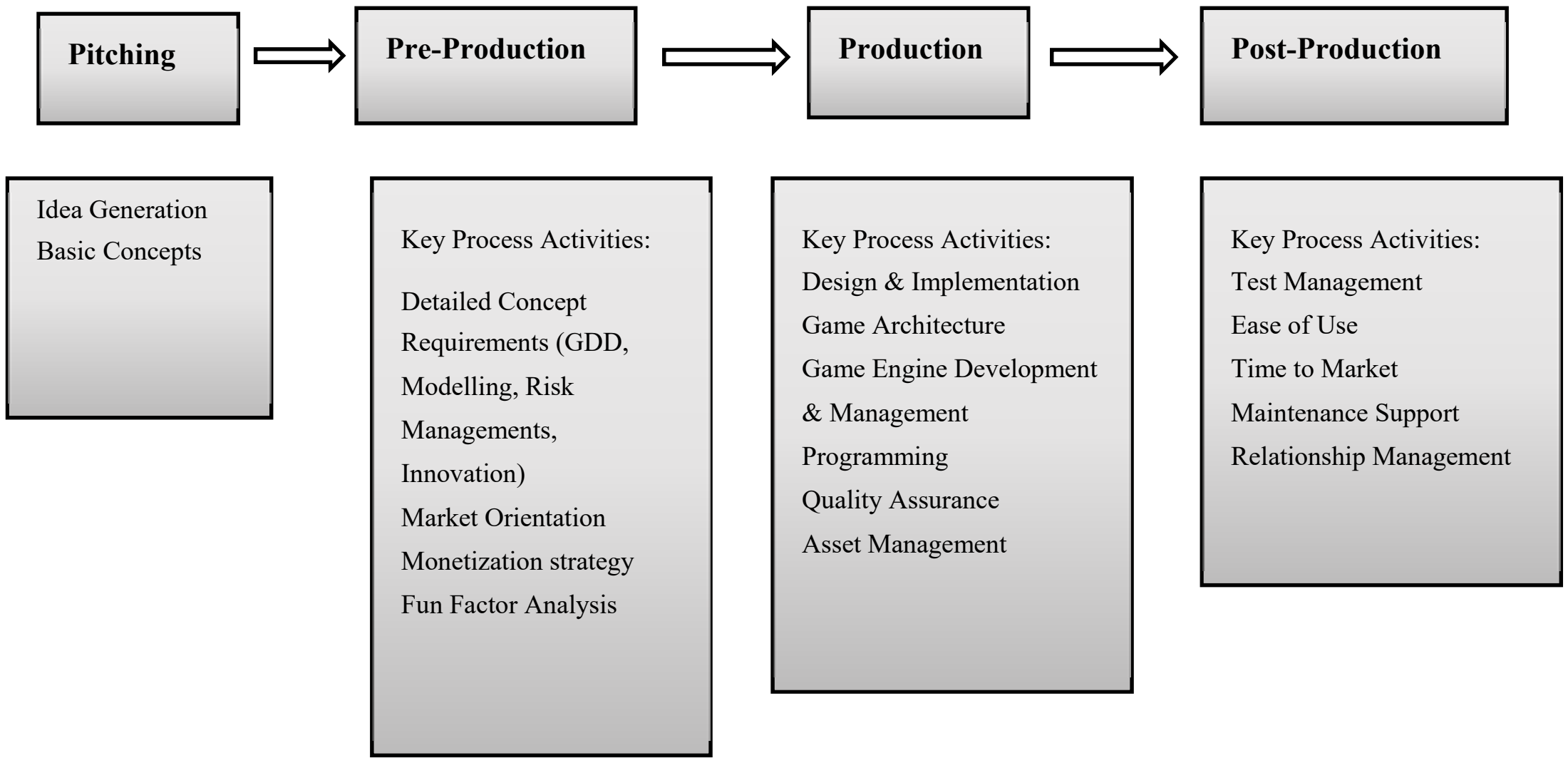

Fig. 1: Game development process.

In traditional software engineering, the development phase usually involves activities like application design and implementation, and the production phase is when the software actually runs and is ready for use. However, in the GDLC, the production phase includes the development process, which is the pre-production phase of the software engineering process, and the production phase of software engineering is actually the post-production phase of the GDLC [32]. Therefore, the GDLC is different from the traditional software engineering process, and many researchers [27] have studied the challenges faced by this domain. The most prominent observation made in these studies is that to address the challenges faced by the GDLC, more rigorous software engineering strategies must be used. However, the proposed GDLCs [6], [29], [30],[32] discussed earlier do not ensure the quality of the development process. Hagan *et al.* [20] published a systematic literature review of software process models used for game development. They concluded that agile and hybrid approaches are used by most organizations for game development. They also reported that Scrum [27], Kanban [33], Rapid Development Application (RAD) [34], XP [35], and incremental [27] methodologies are used by game development organizations. Managing game development has become a much harder process that anyone could have initially imagined, and because of the fragmented nature of the domain, no clear picture of its advancement can be found in the literature.

### *2.2 General Software Development Assessment/Maturity Models*

Humphrey [36] described a software process as a set or order of organizational activities that can be controlled by various entrance and exit criteria imposed by machines, humans, and methods. The actual objective of software process assessment is to develop a high-quality software product within budget and on schedule that meets the needs of its stakeholders. Fuggetta [37] provided a broad description of the software development process, specifying that it should contain software product development, deployment, and maintenance as well as organizational policies and structures, human activities, and the functionalities and technologies used in the process. The software process maturity level of an organization can be assessed by its ability to define, manage, measure, and control the software development process. Assessment of the software process maturity of an organization has emerged as a popular and vital research area in software engineering. The assessment determines the current status of software development processes and has become an essential activity for targeting software process improvement in terms of development and management within the organization. Some well-known international organizations have defined standards for software process assessment, such as the International Standards Organization (ISO), the Software Engineering Institute (SEI), the International Electro-Technical Commission (IEC), and the European Software Institute (ESI).

The popular Capability Maturity Model (CMM) [15] was proposed by the SEI and has been adopted by most organizations in the software industry. The CMM encompasses five maturity levels, ranging from the initial level 1 to the optimizing level

5. Excluding level 1, each level is made up of key process areas (KPAs) and serves as an objective to achieve a certain maturity level. Each KPA has a certain set of features, and if these are collectively achieved, then the goal of the KPA has been accomplished. CMM concepts have also been included in CMMI [16] over time to integrate various disciplines like integrated process and product development, software engineering, system engineering, and supplier sourcing. The BOOTSTRAP [38] methodology has also been used to perform process assessment of organizations by identifying their weaknesses and strengths and offering improvement guidelines. BOOTSTRAP is also made up of five levels but divides the process area into technology, organization, and methodology. Software Process Improvement and Capability dEtermination (SPICE) [39] describes a process assessment reference model for process capability assessment. SPICE is also based on CMM but has six maturity levels with a set of nine documents.

Furthermore, the family of ISO-9000 standards is helpful for setting up a quality management system within an organization for software maintenance and development, as well as for other purposes. ISO-9000-3 can be used to apply ISO-9001 to software supply, development, and maintenance. It also provides guidelines for documentation, responsibility, corrective actions, and software development audits to fulfil ISO-9000 requirements [40]. ISO-12207 [41] provides a framework for improving software engineering and management by grouping broader classes such as primary, support, and organizational activities. ISO/IEC 15504 [42] provides guidance for software process assessment concepts and addresses the two contexts of process capability determination and process improvement. In addition, some of the approaches [43], [44], [45] used for project management maturity models based on CMM have been explored in this context.

All the approaches explored in this proposal concentrate mainly on engineering process assessment specifically for software development activity. As discussed earlier, the digital game development process is different from traditional software development. Its four dimensions have been identified: business, developers, consumers, and the process itself. The software process assessment approaches discussed above cannot be directly used to assess digital game development processes and performance. The proposed approaches and maturity models in the literature capture only the software process aspect of product development. CMMI can be used to assess the process dimension of this research model. Other maturity models have been proposed by various researchers in various domains, such as software product lines [47] and usability of open software systems [46]. These cover the broader aspects of software products by including additional dimensions in the maturity assessment process. To assess digital game maturity, it is necessary to cover, not only the process dimension but also other important dimensions that directly or indirectly contribute to the performance or maturity of digital games in the market. However, for the digital game development process and its performance assessment, no comprehensive method has been proposed that helps an organization identify its strengths and weaknesses in the various activities performed during game development.

### *2.3 Game Development Assessment and Maturity Models*

Digital game process assessment is a very new area of research, and not much work has been reported in this area. Currently, there is no prescribed and systematic way of measuring the maturity level of a digital game development process. According to de Boer *et al.* [48] gamification is the application of game design and game mechanics to motivate and engage people to achieve their targets. They proposed a game maturity model that focusses on using gamification or applied gaming within an organization to gain competitive advantage. The proposed model was based on four perspectives: i) value, ii) process, iii) coverage, and iv) type, with each perspective having five levels. They also analyzed case studies to test their proposed maturity model and demonstrated that the model was an excellent management tool. In fact, the proposed game maturity model did not address assessment of digital game development processes, but rather the use of gamification.

Lee *et al.* [7] examined the ISO 12207 [49] and RUP [50] standards and proposed a game software development process that was applicable to small and medium-sized companies. By conducting panel interviews with practical game developers from the game industry, they identified a set of core elements of game development software and performed requirement analyses for different game genres. The proposed game design process model elaborated inputs and outputs for each activity. The empirical study focussed only on the processes of game development and did not cover broader aspects.

Gorschek *et al.* [51] discussed the process maturity model for market-driven products from the management and requirements engineering perspectives only. The proposed model contained 70 practices, and the interdependencies among them were divided into five process areas. The dependencies among the various practices were defined in the form of AND, OR, REQUIRES, and value-based operators, but they remain to be explored in further detail. Digital games are also a market-driven product, and the proposed model will be helpful in the pre-production phase to address the need to determine and improve process maturity. However, the model is limited to the requirements engineering phase, and its validation remains to be explored. In digital game development field, no studies have been published that directly address the issue of process improvement and assessment. This provided the motivation to propose the Digital Game Maturity Model (DGMM) due to the many challenges faced by organizations in the game development process, including i) lack of research in this area, ii) lack of development processes and good practices, and iii) lack of an assessment model. The following section describes the research methodology used to develop DGMM.

## 3. Research Methodology

The main objective of this research is to propose digital game maturity model that captures different stakeholders' perspectives. In order to propose DGMM, the study is divided into two phases. The first phase involves identification of key factors from three important perspectives and second phase is to propose DGMM based on identified key factors.

**Phase I:** In the first phase, we identified three perspectives that include developers, consumer and business based on literature review. Moreover, identification of key factors for each perspective were also based on available literature in game development domain. An empirical investigation was performed for each perspective in order to identify key factors that have a positive influence or impact on game performance and the development process. For identification of factors from each perspective, research methodology is described below:

*Developer Perspective:* One important game development choice is to consider the developer's perspective to produce good-quality digital games by improving the game development process. To investigate developer's key factor, the research model was developed and its theoretical foundation was based on existing concepts found in the game development literature. Seven key factors were identified and they are: Team configuration and management, Game design document management, Game engine development, Game asset management, Quality of game architecture, Game test management and Programming practices. A quantitative survey questionnaire was developed and conducted to identify key developer factors for an enhanced game development process. For this study, the developed survey was used to test the research model and hypotheses. The results of the study provided the empirical evidence and justification to include factors from the developer's perspective in evaluating the game development process maturity.

*Business Perspective:* Game development organizations are facing high pressure and competition in the digital game industry. Business has become a crucial dimension, especially for game development organizations. The developed research model investigated interrelationship between key business factors and game performance in the market. The model's theoretical foundation is based on a combination of existing concepts found in the game development literature and business models for the game industry. The research model includes seven independent variables: customer satisfaction, market orientation, innovation, relationship management, time to market, monetization strategy, brand name strategy, and one dependent variable i.e. the business performance of the digital game [17]. The results of this study provided evidence that game development organizations must deal with multiple key business factors to remain competitive and handle the high pressure in the digital game industry.

*Consumer Perspective:* Contemporary digital game development companies offer a variety of games for their consumers' diverse tastes. Another important game development choice is considering the consumer perspective to produce quality digital games. The proposed research model was analyzed to find out the associations and interrelationships among the important factors of digital games from a consumer perspective and their influence on digital game performance in the DGI market. The concepts found in the game development literature and in studies, addressing the consumer perspective on digital games provided theoretical foundation for the proposed research model. The research model consists of five independent variables: game engagement, game enjoyment, game characteristics, ease of use, socialization, and one dependent variable, digital game success. A quantitative survey was developed and conducted to identify key consumer factors. For this study, the developed survey was used to test empirically research model and hypotheses. The results provide evidence that game development organizations must deal with multiple key consumer factors to remain competitive and handle high pressure in the digital game industry.

**Phase II:** In the second phase, a *DGMM* is developed by using identified key factors from phase I as a measuring instrument. The maturity and performance of the current game development process can be assessed by using digital game maturity model. The structure of *Digital Game Maturity Model* is composed of assessment frameworks based on identified perspectives such as business, developers, and consumers. The proposed maturity scales include key factors from three perspectives, named as game development process activities (GDPAs). They are used to refer to practices that contribute to the management and development of any game project and listed below with brief description:

1.*Game Design Document Management (GDD):* The GDD has also been identified as an important factor in improving the game development process. The GDD is the outcome of the pre-production phase of game development. It is developed and edited by the game design team to organize their efforts and their development process.

2.*Team Configuration and Management (TCM):* The development of digital games involves multi-disciplinary team configuration and management. Specifically, team configuration and management are considered critical to success of any game development project. Game development requires intensive team management.

3.*Requirements Management and Modelling (RMM):* RMM is also considered important for identification of requirements and helps in development of good GDD.

*4.Game Prototyping (GP)*: Good prototypes are considered as a very important factor in successful game development. Game prototyping for different modules of games is considered as backbone of the successful game project.

*5. Risk Management (Risk_Mgmt.):* Risk management is also considered important in game projects. During the development process, the risk of failing to match the game development strategy was identified as a major cause of problems. A new risk, the "fun factor" was a key element threatening the success of the final game release.

*6.Quality Architecture (QA):* The primary function of the game architecture is to support gameplay. It helps to define challenges by using constraints, concealment, exploration, and obstacles or skill testing. The developer must select the right game architecture for each game project.

*7.Assets Management (AM):* Creation and management of the number of assets required for game development has become challenging. Appropriate mechanisms and strategies are needed to control different versions of assets that are developed for games.

8.*Game Engine and Development (GED)*: A game engine is a software layer that helps in the development process by enabling developers to focus solely on game logic and experimentation (Robin, 2009). Game engines are considered to be a powerful tool by game developers and have been in use for more than two decades.

9.*Test Management (TM):* Game testing is a very important phase of game development. A game can be tested at different levels of development because game testing is different from software testing. Management of game testing during the game development process has clearly come to be of crucial importance for game developers.

10.*Maintenance Support (Mt_S):* Organizations can use appropriate measures to provide maintenance support. To implement better customer service, organizations need to understand their game players, implement player-specific platform services, and take their feedback strongly into consideration.

11.*Fun Factor Analysis (FFA):* Consumers of digital games are motivated to play games because they want to experience fun, and the literature has shown that enjoyment and engagement is a positive reaction of a player during a game play session. The fun factor analysis is important to consider in the game development process.

12.*Ease of Use (EU):* The ease-of-use factor plays a significant role in the game development process. In digital games, ease of use consists of all attributes of the digital game that helps its consumer to control and operate the game easily, within or outside the gameplay.

13.*Market Orientation (MO):* Organization must develops the marketing strategy at beginning of the game development process. Because most of the decisions about game development such as monetization, game design, languages, and demographic locations of game availability will impact the marketing strategy.

14.*Time to Market (TTM):* The TTM approach in a game development organization develops a publishing schedule for the game and provides essential guidelines for development schedules to the developers. The game launch schedule is a crucial business decision that has profound and long-lasting impact on the business performance of an organization in retaining and capturing the market.

15.*Relationship Management (RM):* In successful game development, relationship management plays a significant role. Integrating players in the development process and maintaining excellent working relationships with them helps developers to improve the performance and functionalities of their games.

16.*Monetization Strategy (MS):* For any organization, fulfilment of financial objectives or monetization strategy depends on economically optimizing pricing scheme for the customers, cost structure, and the target customer segment. The impact of monetization can be measured by using overall profitability of the organization as a measure of business performance.

17.*Innovation (I):* Most organizations see innovation in the games as bringing new things to the market and being different from competitors. Innovation in game development can involve application of new ideas at the game level, storyboard production, use of new technology, or creative artistry of the game, with the aim of attracting more gamers and thus creating value in terms of business performance.

18.*Stakeholder Collaboration (SC):* Collaboration between different stakeholders in game development process during each phase is also considered an important aspect of successful development.

The proposed maturity scale includes five levels (in ascending order): Ad-Hoc, Opportunistic, Consistent, Organized, and Optimized. Accordingly, questionnaires were developed for each level, and maturity scale results are presented in the framework described below.

## 4. DGMM for Game Development Organizations

From a game development perspective, development processes must evolve to facilitate new requirements, rapid development, and predictable and repeatable releases. The complex tasks in these development processes require a solid set of best development practices and their exemplary execution.

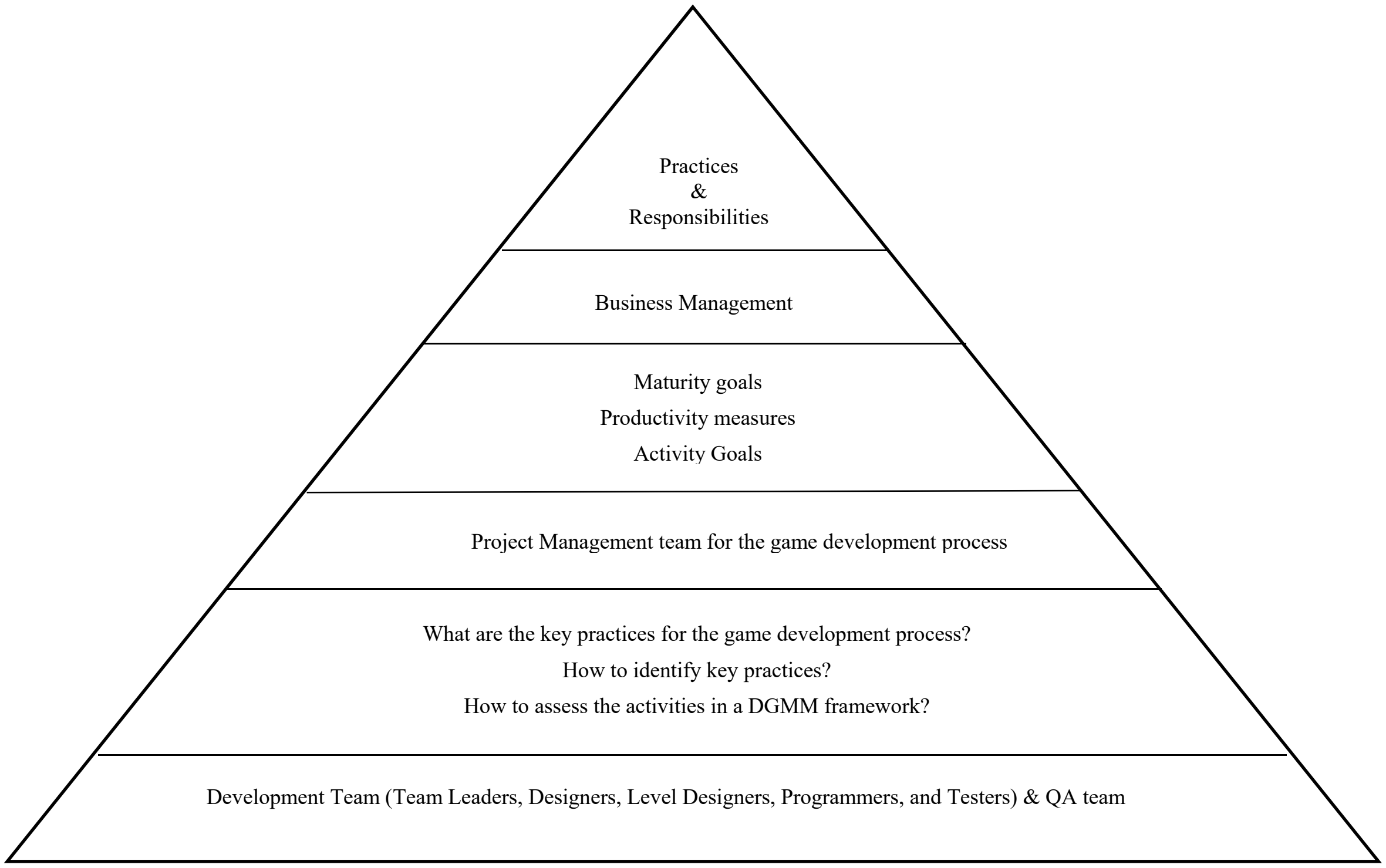

Fig. 2. Three main audiences for a DGMM.

A DGMM aims to establish a comprehensive set of key practices to evaluate digital game development processes. It describes the assessment methodology for development processes and determines the current level of maturity for any game development process in any organization. Furthermore, it is also structured in a way that helps to determine how various key game-development process activities are carried out. The application of DGMM is mainly helpful for three main audiences such as business management, project management, and the development team and the maturity assessment assumes strong coordination among them, as depicted in Fig. 2. Consequently, DGMM is the first study of its kind in the field of game development.

### 4.1 General Scope of a DGMM

The assessment of game development processes is an essential activity to improve current development practices in an organization. Basically, a software engineering maturity model is used for two purposes. First, it provides a strategy to conduct assessment, and second, it provides guidelines to improve current processes. Game development, like any other task, needs some time to show progress or improvements. However, it is not easy to develop an effective and efficient progress plan unless it is based on comprehensive assessment results. Fig.3 represents a comprehensive framework for a game development process assessment exercise for a game development organization. Overall, the game development process involves many key process activities.

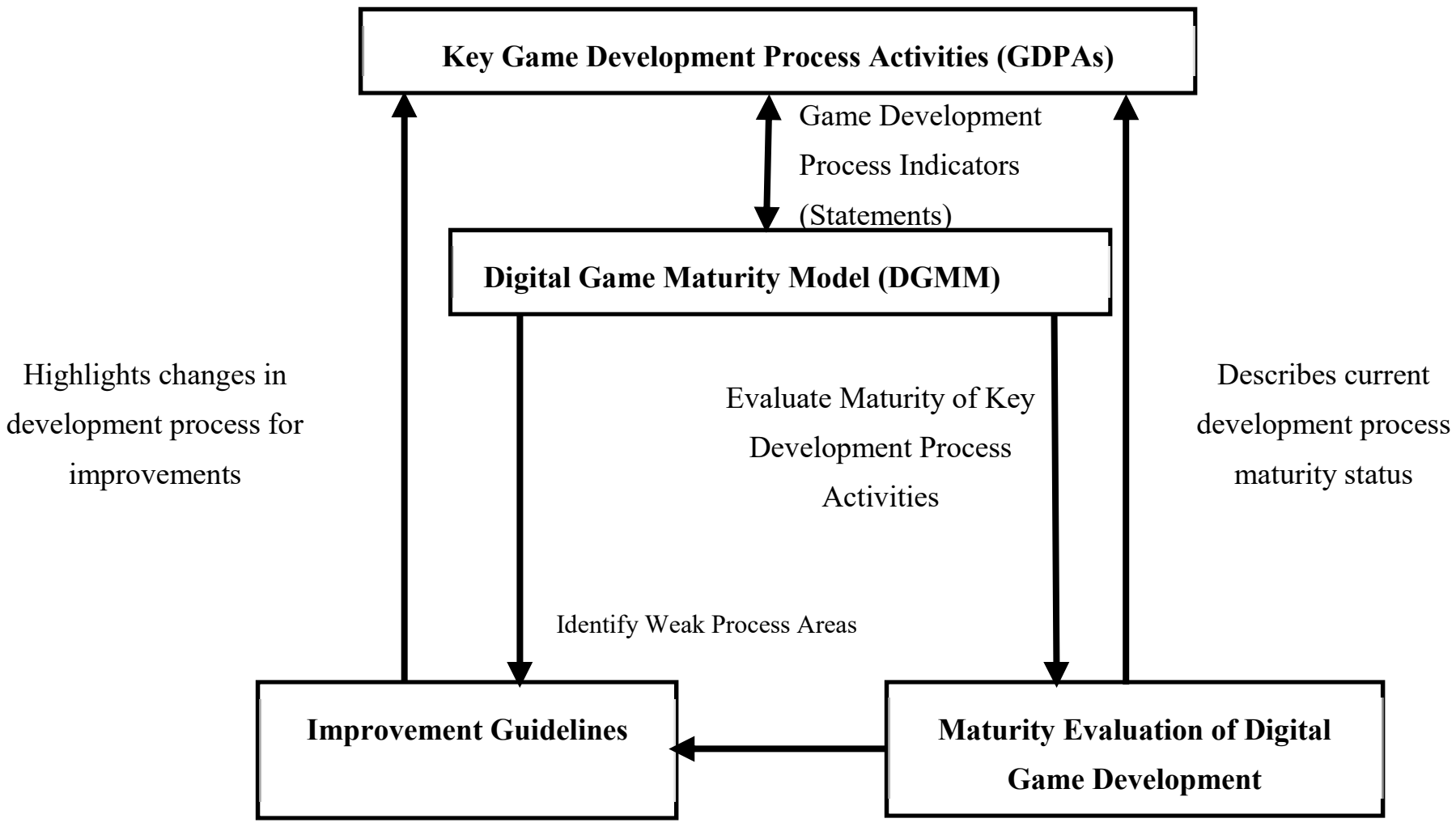

Fig. 3: Scope of a DGMM.

A DGMM presented here uses the key development activities to develop a comprehensive framework consisting of maturity levels and questionnaires to conduct the assessment. Furthermore, a DGMM assesses the current level of game process activities in an organization. The assessment process results in a set of recommendations based on identifying current process weaknesses that need improvement. However, the proposed DGMM does not provide any guidelines for current process improvements, which may be considered as a future project.

### 4.2 Configuration of a DGMM

The functional configuration of a DGMM consists of eighteen key process activities, i.e., activities essential for the game development process. Specifically, Table 1 describes the domains and hierarchy of a DGMM. The eighteen key development process activities used in this model, as mentioned above, are divided into four dimensions: game design strategy, game development methodology, game playability and usability, and finally the business performance dimension.

The Game Design Strategy dimension mainly covers GDD Management, TCM, RMM, GP, and Risk_Mgmt. The Game Development Methodology dimension includes three important process activities, QA, AM, and GED. Authors conducted an empirical study from a developer's perspective of key game development practices and selected certain key practices for the Game Design Strategy and Game Development Methodology dimensions that had been found important for game development. The Game Playability & Usability dimension covers TM, Mt_S, FFA, and EU. Authors also performed an empirical investigation from a consumer's perspective of successful game development factors and selected key practices for the Game Playability and Usability dimension that have a positive impact on digital game success.

Table 1: Configuration of a DGMM.

| Dimensions | Activity ID (AID) | Game Development Process Activities (GDPA) |
|---|---|---|
| Game Design Strategy | 1 | Game design document management (GDD) |
| | 2 | Team configuration and management (TCM) |
| | 3 | Requirements management and modelling (RMM) |
| | 4 | Game prototyping (GP) |
| | 5 | Risk management (Risk_Mgmt.) |
| Game Development Methodology | 6 | Quality architecture (QA) |
| | 7 | Assets management (AM) |
| | 8 | Game engine and development (GED) |
| Game Playability & Usability | 9 | Test management (TM) |
| | 10 | Maintenance support (Mt._S) |
| | 11 | Fun factor analysis (FFA) |
| | 12 | Ease of use (EU) |
| Business Performance | 13 | Market orientation (MO) |
| | 14 | Time to market (TTM) |
| | 15 | Relationship management (RM) |
| | 16 | Monetization strategy (MS) |
| | 17 | Innovation (I) |
| | 18 | Stakeholder collaboration (SC) |

The Business Performance dimension of a DGMM includes MO, TTM, RM, MS, I, and SC. Aleem *et al.* [17] investigated key business factors for game performance in the DG market. The selected key practices in Business Performance dimension are important key factors for the business performance of any game. The empirical studies carried out to capture various perspectives on the key factors in the game development process [17] and their presence in the literature provide the motivation to include the eighteen key process activities under the four DGMM dimensions. These eighteen important key practices are the foundation of the assessment questionnaires, which consist of “statements”. These statements describe the effectiveness of the particular activities as they contribute to game development and management.

### 4.3 Framework of a DGMM

To define maturity level, ranking is considered an important part of the process assessment methodology. As discussed earlier, many software process assessment models such as CMMI [16], SPICE [39], and BOOTSTRAP [38] use ranking to define their proposed maturity models. The proposed DGMM also uses ranking to define the level of maturity. In ascending order, these levels are: *Ad-Hoc, Opportunistic, Consistent, Organized,* and *Optimized*. An assessment questionnaire is developed for each maturity level. For each level, the questionnaire contains a number of statements that are divided into eight GDPAs. The maturity level of the game development process within an organization is determined by the extent to which the three audiences identified earlier agree with each statement in the questionnaire. Assessment questionnaires for each maturity level in this study are designed and written specifically for a DGMM.

The methodology for assessing the current GDPA aims to establish a comprehensive strategy for evaluating the current GDPA maturity level in a game project. Furthermore, it is designed to identify the systematic way in which various GDPAs are performed during the game development project life cycle. In general, a DGMM attempts to coordinate the game development process with identified GDPAs. Table 2 shows a DGMM framework. Each level of maturity is defined by a set of statements that cover all eighteen GDPAs used in this study. The total number of statements differs for each maturity level and GDPA. Abbreviations will be used for the GDPAs in the rest of the paper. These includes Game Design Document Management (GDD), Team Configuration and Management (TCM), Requirements Management and Modelling (RMM), Game Prototyping (GP), Risk Management (Risk_Mgmt.), Quality Architecture (QA), Assets Management (AM), Game Engine and Development (GED), Test Management (TM), Maintenance Support (Mt._S), Fun Factor Analysis, (FFA), Ease of Use (EU), Market Orientation (MO), Time to Market (TTM), Relationship Management (RM), Monetization Strategy (MS), Innovation (I), and Stakeholder Collaboration (SC). The characteristics of game development organizations are described in the following sub-section. Specifically, each game development organization will be described in terms of the GDPA maturity scale and the measuring instrument designed for a DGMM. A DGMM measuring instrument uses the following symbols and abbreviations:

GDPA = Game Development Process Activity,
ML = Maturity Level (an integer),
S = Statement,
GN = Game development activity Number (integer),
SN = Statement Number (an integer).

Table 2: DGMM framework.

| **Maturity level** | **GDPAs and number of statements in assessment questionnaire** | | | | | | | | | | | | | | | | | |
|---|---|---|---|---|---|---|---|---|---|---|---|---|---|---|---|---|---|---|
| | GDD | TCM | RMM | GP | Risk_ Mgmt | QA | AM | GED | TM | Mt._S | FFA | EU | MO | TTM | RM | MS | I | SC | Total |
| **Ad-Hoc** | 2 | 1 | 3 | 1 | 1 | 2 | 1 | 2 | 2 | 1 | 2 | 2 | 2 | 2 | 3 | 2 | 1 | 1 | 31 |
| **Opportunistic** | 5 | 2 | 4 | 2 | 2 | 3 | 4 | 3 | 3 | 2 | 3 | 3 | 3 | 3 | 3 | 2 | 3 | 1 | 51 |
| **Consistent** | 4 | 3 | 4 | 2 | 3 | 4 | 3 | 3 | 5 | 2 | 2 | 3 | 4 | 2 | 3 | 3 | 3 | 1 | 54 |
| **Organized** | 4 | 4 | 3 | 4 | 3 | 4 | 2 | 3 | 4 | 2 | 4 | 3 | 4 | 2 | 2 | 3 | 2 | 1 | 54 |
| **Optimized** | 4 | 2 | 3 | 3 | 3 | 3 | 2 | 2 | 3 | 2 | 2 | 2 | 3 | 2 | 2 | 3 | 1 | 1 | 53 |

### *4.3.1 Ad-Hoc (Level I)*

The "*Ad-Hoc*" level is the initial level of a DGMM, which indicates that the game development organization does not have an organized and stable methodology for game development. If an organization is at the "*Ad-Hoc*" level, there is a lack of understanding of best practices for game development. Furthermore, there is no evidence that the organization develops games using specified software engineering practices or performs various development-related activities in a co-ordinated way. Instead, the organization develops different games independently and performs their development activities on an *Ad-Hoc* basis. Moreover, there is no protocol established to reuse assets for other game projects, nor is there evidence of following requirements and management strategy or a game development methodology. The organization does not perform any assessment of game playability and usability, nor does it have the technical resources and skills to manage game development projects properly. The following assessment questionnaire shows the GDPA maturity of a game development organization at Level 1 in terms of key GDPAs.

GDPA 1.1 GDD Management
S.1.1.1 There is no evidence of developing a GDD.
S.1.1.2 The requirements gathering process is *Ad-hoc*, and the skills and resources to develop a GDD are absent.

GDPA 1.2 Team Configuration & Management
S.1.2.1 There is no established team configuration and management strategy.

GDPA 1.3 Requirements Modelling and Management
S.1.3.1 The development team is not using any specific notation language to model game requirements.
S.1.3.2 There is a lack of knowledge of requirements modelling.
S.1.3.3 No market analysis is performed to gather requirements.

GDPA1.4 Game Prototyping
S.1.4.1 No prototype is developed at the end of the pre-production phase.

GDPA 1.5 Risk Management
S.1.5.1 No risk management is performed at any phase of game development.

GDPA 1.6 Quality of Architecture

S.1.6.1 The management team lacks an understanding of game architecture evaluation attributes.

S.1.6.2 There is no evidence that the development team performs a systematic architecture evaluation.

GDPA 1.7 Asset Management

S.1.7.1 There is no evidence of any planned asset management for the various game assets.

GDPA 1.8 Game Engine Development & Management

S.1.8.1 The development team is totally dependent on commercially available game engines.

S1.8.2 The development team does not have enough skills and technical knowledge to manage or develop a game engine.

GDPA 1.9 Test Management

S.1.9.1 There is no proper plan for game testing.

S.1.9.2 No testing is performed to ensure adherence to functional and non-functional requirements.

GDPA 1.10 Maintenance Support

S.1.10.1 No maintenance support is provided by the development team to its customers.

GDPA 1.11 Fun Factor Analysis

S.1.11.1 The game engagement and enjoyment factor is not considered important for game success.

S.1.11.2 No market analysis is performed to enhance the fun factor in games.

GDPA 1.12 Ease of Use

S.1.12.1 There is no proper plan to analyze the usability factor.

S.1.12.2 The development team does not have enough knowledge and skills to improve game usability.

GDPA 1.13 Market Orientation

S.1.13.1 The project team does not consider that market orientation is an important factor in the game business.

S.1.13.2 There is no evidence that the development team performs market analysis of the game type and the target audience.

GDPA 1.14 Time to Market

S.1.14.1 The time to market factor is not considered important for game launch.

S.1.14.2 The development team does not conduct any market reviews to update game publishing time.

GDPA 1.15 Relationship Management

S.1.15.1 The developed game has complex gameplay and goals are not clearly defined.

S.1.15.2 There is no evidence of feedback mechanisms.

S.1.15.2 The developed game has too many switching players and cannot retain players for a long time.

GDPA 1.16 Monetization Strategy

S.1.16.1 There is no evidence that management has developed a monetization strategy.

S.1.16.2 The game revenue model is not successful in convincing players to buy virtual assets.

GDPA 1.17 Innovation

S.1.17.1 There is no evidence of any research and development component.

GDPA 1.18 Stakeholder Collaboration

S.1.18.1 No collaboration exists among stakeholders for game development related decisions.

### *4.3.2 Opportunistic (Level II)*

The next DGMM maturity level has been defined as “Opportunistic”. At this level, the management and development teams realize the importance of best practices related to game development and show interest in adopting them. In addition, the management team makes efforts to collect data for requirements analysis, playability, and usability factors on an occasional basis. The development team is also interested in acquiring knowledge and skills related to appropriate development methodologies. Both the management and development teams agree that a proper game assessment strategy for game playability and usability is important and also recognize the importance of assessing current practices. However, the organization lacks systematic planning and strategy for its revenue model and market analysis. The organization does not maintain any kind of documentation related to the pre-production, production, and post-production phases of game development. Moreover, there are no clear guidelines for relationship management and team collaboration. At this earlier stage of a DGMM, an organization is concentrating on understanding how to develop quality games that will be successful and attract more consumers. This is why this stage has been called “opportunistic”: an organization at this level sees the opportunity to build its understanding about best practices and to acquire enough resources and skills to move to the next level. Overall, an organization at this level understands the importance of adopting best practices for game development and is in process of establishing defined protocols for GDPAs. The following set of statements must be satisfied by an organization in Level 2.

GDPA 2.1 GDD Management

S.2.1.1 Management believes that a well-defined game design document is helpful in the game production phase.

S.2.1.2 The project team agrees to follow design principles for gameplay, mechanics, and documentation.

S.2.1.3 Game designers believe that dictionaries of design terms are important in transforming the GDD from pre-production to the production phase.

S.2.1.4 The technical and creative team agrees that a vocabulary of game design terminology and the development of visual languages for design modelling are important in the pre-production phase.

S.2.1.5 No formal game design document is developed at the end of the pre-production phase.

GDPA 2.2 Team Configuration & Management

S.2.2.1 The project manager and the team lead agree that a development team organized by discipline can perform its assigned tasks more effectively than one organized by features.

S.2.2.2 There is no formal protocol established for collaboration among development team members.

GDPA 2.3 Requirements Modelling and Management

S.2.3.1 Project managers and team leaders understand that requirements modelling helps in understanding game development requirements.

S.2.3.2 The development team is making efforts to acquire technical knowledge and understanding to develop media design documents, game feature plans, and technical design documents.

S.2.3.3 The development team is committed to analyzing market data for new game trends and consumer requirements.

GDPA2.4 Game Prototyping

S.2.4.1 The development team is committed to following a formal prototyping method because it is considered crucial for successful game development.

S.2.4.2 The development team believes that a high-fidelity game prototype provides a higher degree of sophistication and is therefore, closer to the final product than a lower-fidelity prototype.

GDPA 2.5 Risk Management

S.2.5.1 The project manager believes that risks related to game development must be identified in the pre-production phase.

S.2.5.2 The management team is committed to acquiring knowledge and resources related to game development risk identification strategy.

GDPA 2.6 Quality of Architecture

S.2.6.1 The development team is acquiring knowledge and skills to model game architecture properly.

S.2.6.2 The development team is committed to establishing clear guidelines and methodology for game architecture.

S.2.6.3 Quality and functional attributes are not well defined.

GDPA 2.7 Asset Management

S.2.7.1 The management team is committed to developing a proper strategy for game asset creation and management.

S.2.7.2 The project manager believes that realism and performance analysis must be part of the asset creation process.

S.2.7.3 The project manager agrees that realism and control investigation of assets must be performed before asset creation.

S.2.7.4 There is need of proper asset management tool to manage various game assets.

GDPA 2.8 Game Engine Development & Management

S.2.8.1 The development team has adequate resources and skills to use and manage game engine modules.

S.2.8.2 The project manager is committed to providing training to the development team for game engine development.

S.2.8.3 The development team uses commercially available game engines to develop games.

GDPA 2.9 Test Management

S.2.9.1 The project team recognizes the need for a game testing plan during the pre-production phase of game development.

S.2.9.2 The project team collects information on how to validate game functional and non-functional requirements through testing.

S.2.9.3 The project manager agrees that test management will give insights into how players play the game and the pros and cons of the game design and will finally be helpful in making a complete, balanced, and fun to play game.

GDPA 2.10 Maintenance Support

S.10.1 The development team is working on developing a forum where game consumers can report playability, bugs, errors, and other game-related issues.

GDPA 2.11 Fun Factor Analysis

S.2.11.1 The project team collects data on how to enhance the enjoyment and engagement factor for their games.

S.2.11.2 The management team promotes innovative ideas to develop games that provide many stimuli from different sources with interesting and attractive gameplay.

GDPA 2.12 Ease of Use

S.2.12.1 The members of the development team are acquiring knowledge and skills to improve ease of use for digital games.

S.2.12.2 The project manager agrees that ease of use is a fundamental driver for the commercial success of digital games.

S.2.12.3 There is a lack of systematic strategy to enhance the level of usability in digital games.

GDPA 2.13 Market Orientation

S.2.13.1 No formal strategy has yet been developed to perform detailed market analysis of game types and their target audience.

S.2.13.2 There are lack of resources to perform market analysis.

S.2.13.3 No well-defined communication protocol exists for information sharing.

GDPA 2.14 Time to Market

S.2.14.1 The development team occasionally studies and researches development updates.

S.2.14.2 Game publishing is not influenced by the time to market factor.

S.2.14.3 The project team agrees that time to market is important for game publishing.

GDPA 2.15 Relationship Management

S.2.15.1 No formal player integration strategy has been established for game development.

S.2.15.2 Feedback mechanisms have been developed to resolve player concerns and issues.

S.2.15.3 Customer profiling is performed on an occasional basis.

GDPA 2.16 Monetization Strategy

S.2.16.1 The revenue model is not well defined.

S.2.16.2 No progressive growth has been observed in the last two years.

GDPA 2.17 Innovation

S.2.17.1 No well-defined policy for research and development (R&D) has been established.

S.2.17.2 Innovative ideas are considered important for new game development projects.

S.2.17.3 The development team occasionally studies and reviews development updates and searches for innovative ideas.

GDPA 2.18 Stakeholder Collaboration

S.2.18.1 Project managers and development team agrees that collaboration among all stakeholders is important to identify game requirements completely.

#### *4.3.3 Consistent (Level III)*

An organization at level III is consistently trying to define policies and strategies for game development projects. Moreover, the organization is able to establish an infrastructure for game development projects by completing identified GDPAs. Interest in developing a strategic plan shows that the organization is committed to developing good-quality products and trying to address the challenges faced by the game development team. An organization that can develop a strategy for game design documents, establish protocols, and acquire enough resources and skills for requirements modelling and management is exhibiting sufficient knowledge of the domain. Subsequently, such an organization in the production phase is committed to developing a game architecture that fulfils quality attributes and is trying to manage game assets effectively. To ensure the required game playability and usability, game testing, fun factor analysis and maintenance support are considered mandatory GDPAs by the organization. The organization is trying to develop business strategies and documentation and in process of establishing clear guidelines for carrying out the game development process. Accordingly, the management and development teams have acquired enough training in development methodologies. Overall, the organization is able to understand game requirements and development methodologies, the playability and usability factors, and the digital game business performance indicators. Accordingly, it is trying to be consistent in its GDPAs for the development process. In the following measuring instrument, the set of statements describes the level of maturity of an organization at level III.

GDPA 3.1 GDD Management

S.3.1.1 The design team is committed to and in the process of developing dictionaries for design terms.

S.3.1.2 The design team is developing design guidelines and concepts.

S3.1.3 The development team is acquiring knowledge about design principles for gameplay, mechanics, and documentation.

S.3.1.4 The design team is in the process of developing a strategy to produce a GDD.

GDPA 3.2 Team Configuration & Management

S.3.2.1 Sub-teams are organized on a discipline basis.

S3.2.2 There is an established protocol for collaboration among development team members.

S.3.2.3 The team leader is committed to involving all team members in prioritizing the various tasks for each sprint or milestone.

GDPA 3.3 Requirements Modelling and Management

S.3.3.1 The development team has the required resources and technical knowledge to model and manage requirements.

S.3.3.2 The development team is able to develop a strategy for producing the media design, feature plan, and technical design document.

S3.3.3 The requirements document clearly identifies the structural layout of the game architecture.

S.3.3.4 Team leaders collect and analyze data from the market and gather requirements for customer profiling on a regular basis.

GDPA3.4 Game Prototyping

S.3.4.1 The development team has developed multiple prototypes because games are highly visual, functional, and interactive applications. Hence, they believe that a single prototype is in most cases insufficient to capture all aspects of a game.

S.3.4.2 The development team is acquiring resources and skills to develop a proper strategy for game prototyping.

GDPA 3.5 Risk Management

S.3.5.1 Risks related to game usability and playability, supportability, performance, budgeting, and scheduling are identified during the pre-production phase.

S.3.5.2 Strategies are developed to manage risks related to development strategy, staffing, budgeting, scheduling, inadequate specifications, and the fun factor of games.

S.3.5.3 The management team is able to perform reactive risk management.

GDPA 3.6 Quality of Architecture

S.3.6.1 The project manager is in process of establishing clear guidelines and a well-documented methodology for game architecture.

S.3.6.2 The management team has acquired sufficient technical knowledge to develop and evaluate game architecture.

S.3.6.3 Game prototyping is used to analyze the interconnection among the various gameplay modules.

S.3.6.4 The management team is committed to using best software engineering practices to evaluate game architecture.

GDPA 3.7 Asset Management

S.3.7.1 There is a consistent strategy for game asset management.

S.3.7.2 An asset management system has been implemented for storing and managing all game assets.

S.3.7.3 The project team has adequate resources and skills to analyze any new asset management tool that is introduced and to acquire it if it provides better asset management.

GDPA 3.8 Game Engine Development & Management

S.3.8.1 The selected game engine is able to handle diverse type of input and output.

S.3.8.2 The game engine is able to provide resource and asset management.

S.3.8.3 Integration of all technological aspects is done easily by the development team.

GDPA 3.9 Test Management

S.3.9.1 A game testing plan is established and well documented during the pre-production phase.

S.3.9.2 Internal testers have acquired sufficient knowledge to assess functional, playability, and usability requirements.

S.3.9.3 External testers participate in game testing to identify the playability and usability of specific game projects.

S.3.9.4 Game testing has provided evidence to remove unsuccessful parts of a game design.

S.3.9.5 Testing is usually started during the pre-production phase to avoid later-stage modifications.

GDPA 3.10 Maintenance Support

S.3.10.1 The project team is committed to improve maintenance support for developed games.

S.3.10.2 A log has been maintained regarding issues faced by game consumers who report errors or bugs in a purchased game.

GDPA 3.11 Fun Factor Analysis

S.3.11.1 A strategic plan has been defined to gather consumer requirements and perform market analysis to enhance the consumer playing experience in term of game workload, rewards, full control, skill level, and storyline.

S.3.11.2 The project team is in process of defining metrics to perform fun factor analysis.

GDPA 3.12 Ease of Use

S.3.12.1 There is a well-defined strategy and clear guidelines for developing game tutorials.

S.3.12.2 The development team includes game tutorials to provide internal and external consistency to game consumers.

S.3.12.3 The project team is in process of defining metrics to measure ease of use.

GDPA 3.13 Market Orientation

S.3.13.1 Market analysis is performed occasionally.

S.3.13.2 A communication protocol for dissemination of market intelligence has been defined.

S.3.13.3 A market orientation strategy is developed during the pre-production phase.

S.3.13.4 Game concepts are influenced by competitors.

GDPA 3.14 Time to Market

S.3.14.1 The management team performs regular market reviews and development updates.

S.3.14.2 Development schedules are adjustable based on market updates.

GDPA 3.15 Relationship Management

S.3.15.1 The development team participates in online game communities to support and identify player issues.

S.3.15.2 Management is trying to define player integration strategies for game development.

S.3.15.3 Data mining techniques are used to extract, manipulate, and produce data quickly for consumer profiling.

GDPA 3.16 Monetization Strategy

S.3.16.1 A well-defined revenue model has been developed.

S.3.16.2 Sales revenue has been growing over a time period.
S.3.16.3 The defined model is able to reduce debt.

GDPA 3.17 Innovation
S.3.17.1 Management believes that R&D investment yields positive results in the near future.
S.3.17.2 Management is in process of defining an R&D policy.
S.3.17.3 The development team use innovative ideas successfully for development and game level repositioning.

GDPA 3.18 Stakeholder Collaboration
S.3.18.1 Collaboration among all stakeholders is performed on an occasional basis.

#### *4.3.4 Organized/Predictable (Level IV)*

The fourth level of a DGMM is referred to as "organized and predictable". An organization is considered to be at this level if it has been successful in developing well-defined guidelines for game development activities and the project team has acquired all resources and technical skills to address issues in the game development process. The management team is able to develop requirements models that help to visualize the interconnections among the various game modules, player interaction patterns, procedures, rules, resources, conflicts, boundaries, outcomes, rewards, and goals. The members of multidisciplinary teams can collaborate and identify bottlenecks in the development phases. Moreover, the management team takes a proactive stance with regard to risk management and innovation. Once developed, game projects can retain and satisfy their customers, and their revenue model fits into the organization financial model. Market analysis is performed by the organization on a regular basis for the time to market factor. Furthermore, defined metrics are used to analyze the fun factor and ease of use in games. Proper test and asset management strategies are in place. Overall, the GDPAs in such an organizations are streamlined, quantifiable, and well documented for any game project, and is, considered to be at level IV of a DGMM. The resulting set of statements listed below applies to an organization at level IV.

GDPA 4.1 GDD Management
S.4.1.1 The development team has adequate resources to develop the GDD.
S.4.1.2 The GDD offers clear guidelines for the transformation from pre-production to the production phase.
S.4.1.3 Dictionaries for design terms serve as a basis for communication among professionals and for project documents.
S.4.1.4 The development team is following proper game design guidelines and benchmarking existing ones.

GDPA 4.2 Team Configuration & Management
S.4.2.1 After each milestone, team members meet and discuss the progress of each project on regular basis.
S.4.2.2 Team management helps in identifying topics of interest and production bottlenecks effectively.
S.4.2.3 All stakeholders are involved in the decision process for any significant change in game design or architecture during the production phase.
S.4.2.4 The development plan is well documented and communicated to all team members.

GDPA 4.3 Requirements Modelling and Management
S.4.3.1 The requirements document for the game covers the scope of the final game.
S.4.3.2 Game requirements are well documented and clearly identified.
S.4.3.3 The requirements model helps to visualize the interconnections among the various game modules, player interaction patterns, procedures, rules, resources, conflicts, boundaries, outcomes, rewards, and goals.

GDPA4.4 Game Prototyping
S.4.4.1 The selected prototyping tool provides flexibility and stability to make adjustments after feedback.
S.4.4.2 The development team follows a software prototyping lifecycle that includes requirements identification for the final models, textures, particle systems, materials, level geometry and lighting, audio, and animation.
S.4.4.3 The prototyping strategy helps to identify various options for balancing game mechanics and aesthetics.
S.4.4.4 Game prototyping provides an early insight into how the final game will be played.

GDPA 4.5 Risk Management
S.4.5.1 Risk assessment is considered mandatory during the pre-production phase of game development.
S.4.5.2 Risk-related tasks are clearly identified by the management team for each milestone.
S.4.5.3 Management is able to perform proactive risk management.

GDPA 4.6 Quality of Architecture
S.4.6.1 Game architecture quality attributes such as performance, correctness, usability, testability, security, and scalability are well defined and documented.
S.4.6.2 The development team is using specific defined qualitative metrics to measure the quality of gameplay.
S.4.6.3 Gameplay is divided into different modules, and there is a separation of concerns because they can be modified separately without impacting other modules.
S.4.6.4 Gameplay modules are extensible because they can be plugged into other game projects.

GDPA 4.7 Asset Management
S.4.7.1 The asset management system group different assets, provide version control and simplify workflow.
S.4.7.2 The asset management system can handle any size of files for graphics, video, and sound.

GDPA 4.8 Game Engine Development & Management
S.4.8.1 Multiplatform development is supported by the selected game engine.
S.4.8.2 The game engine is capable of integrating other embedded tools to enhance or extend its current capabilities.
S.4.8.3 The selected development tool enables developers to manage, visualize, and maintain transformations so they can be helpful in game development.

GDPA 4.9 Test Management
S.4.9.1 The project has a defined roadmap for testing during each phase of game development to test different game modules.
S.4.9.2 The game is tested for performance under various loads.
S.4.9.3 The testing unit learns from previous game testing experiences and avoids repeating the same mistakes.
S.4.9.4 A well-established game testing management plan with quantifiable metrics has been implemented to perform testing regularly.

GDPA 4.10 Maintenance Support
S.4.10.1 The development team has developed an appropriate maintenance support unit for consumers.
S.4.10.2 The project team monitors the maintenance support unit for effective and efficient consumer support.

GDPA 4.11 Fun Factor Analysis
S.4.11.1 A strategic plan has been fully implemented to enhance the consumer play experience.
S.4.11.2 The project team regularly monitors the outcome of innovative ideas for consumer engagement and enjoyment.
S.4.11.3 Consumer feedback is regularly collected with regard to their gameplay experience, to enhance the market presence of the game.
S.4.11.4 Management team uses defined metrics for fun factor analysis.

GDPA 4.12 Ease of Use

S.4.12.1 The project team analyzes consumer feedback to make sure that consumers can easily manipulate the game controls to take actions which help in achieving game goals.

S.4.12.2 Consumer feedback about game ease of use is regularly collected and maintained.

S.4.12.3 The project team is using well-defined metrics to measure ease of use in particular games.

GDPA 4.13 Market Orientation

S.4.13.1 Market analysis is performed on a regular basis to identify target audiences and in-demand games.

S.4.13.2 New projects are in line with consumer requirements.

S.4.13.3 Competitor analysis is performed to develop new market plans.

S.4.13.4 Skills and resources are adequate to perform market analysis.

GDPA 4.14 Time to Market

S.4.14.1 Games are launched in response to competitor actions.

S.4.14.2 The timing of game launch helps in increasing market presence.

GDPA 4.15 Relationship Management

S.4.15.1 The organization performs consumer profiling for profitability analysis and retention modelling on a regular basis.

S.4.15.2 Developed games are able to retain their existing consumers and attract new ones.

GDPA 4.16 Monetization Strategy

S.4.16.1 Developed games are able to acquire more consumers for less investment.

S.4.16.2 To attract new consumers, cross-platform offerings are in place.

S.4.16.3 The revenue model for developed games fits into the financial model of the organization.

GDPA 4.17 Innovation

S.4.17.1 Reactive and proactive innovation measures for game development are supported by management.

S.4.17.2 An R&D roadmap is successfully used for game development.

GDPA 4.18 Stakeholder Collaboration

S.4.18.1 All stakeholders collaborate on a regular basis, but it is not considered mandatory to involve them in game development-related decisions.

#### *4.3.5 Optimized (Level V)*

The highest level of a DGMM is referred to as "optimized". At this level, the GDPAs play an important role in the business performance of the games developed by the organization. There is strong evidence that management and development teams collaborate closely to manage and develop games effectively. Stakeholders are involved in all game development-related decisions. The organization learns from its past game development experiences and on this basis is in process of optimizing its current GDPAs. Hence, training and attaining knowledge about new technologies and skills related to game development is a continuous process in the organization. Game requirements are reviewed and revised on a regular basis when required. The development team has adequate resources and skills to develop its own game engines for game development or to enhance the capabilities of existing ones by adding middleware. A developed testing plan is able to keep track of functional and non-functional requirements test outcomes and uses the results to improve game quality and playability. A blend of playability and usability methods in addition to innovative ideas are used to enhance the consumer playability experience in term of challenges, storyline, game-level curiosity, full control, and feeling of independence. Moreover, the revenue model contributes to strengthening the financial position of the organization. The measuring instrument for a DGMM at level V is illustrated below.

GDPA 5.1 GDD Management

S.5.1.1 Defined game design guidelines and concepts are followed for all new game development projects.

S.5.1.2 The GDD is well understandable by all stakeholders.

S.5.1.3 The GDD is available to all development team members at the beginning of the production phase.

S.5.1.4 A log is maintained to record development team members' complaints regarding GDD transformation issues.

GDPA 5.2 Team Configuration & Management

S.5.2.1 Team configuration and management demonstrate a positive impact on game development activities.

S.5.2.2 Team members are satisfied with the communication and collaboration protocol.

GDPA 5.3 Requirement Modelling and Management

S.5.3.1 The target market segment is fully captured by the identified requirements of a particular game.

S.5.3.2 Game requirements are reviewed and revised on a regular basis when required.

S.5.3.3 The quality attribute of games is accommodated by identified requirements.

GDPA5.4 Game Prototyping

S.5.4.1 Prototyping helps in improving and developing the final game efficiently.

S.5.4.2 Prototyping helps in identifying game mechanics, rules, and algorithms.

S.5.4.3 The developed prototype refines the created content of the game and also balances the gameplay.

GDPA 5.5 Risk Management

S.5.5.1 Risk assessment is helpful in reducing associated development risks.

S.5.5.2 There is a backup plan to handle identified risks and explore other solutions that would reduce or eliminate risk.

S.5.5.3 The development team always has a functional and technical design specification with a complete risk assessment document before the start of the production phase for all projects.

GDPA 5.6 Quality of Architecture

S.5.6.1 The management team is continuously improving the evaluation process for game architecture quality.

S.5.6.2 Game architecture documents are reviewed and updated regularly to avoid future bottlenecks.

S.5.6.3 Game architecture includes robustness features that enable the game to be functional in unexpected circumstances.

GDPA 5.7 Asset Management

S.5.7.1 The asset management system can reduce duplication of assets and remove outdated assets from the asset library.

S.5.7.2 Assets created for a game fit into the game concept and have a positive effect on game appearance.

GDPA 5.8 Game Engine Development & Management

S.5.8.1 The development team has adequate resources and skills to develop its own game engines for game development or to enhance the capabilities of existing ones by adding middleware.

S.5.8.2 Game engines are reused for different game projects.

GDPA 5.9 Test Management

S.5.9.1 The selected testing approach ensures game performance and quality.

S.5.9.2 The testing team experiments with innovative techniques on a regular basis to improve the game testing process.

S.5.9.3 A developed test plan keeps track of functional and non-functional requirements test outcomes and uses the results to improve game quality and playability.

GDPA 5.10 Maintenance Support

S.5.10.1 The maintenance support system team regularly examines, maintains, and improves the support system for effective and easy reporting service.

S.5.10.2 The project team is continuously improving the maintenance support system for developed games.

GDPA 5.11 Fun Factor Analysis

S.5.11.1 A blend of playability and usability methods in addition to innovative ideas are used to enhance the consumer playability experience in term of challenges, storyline, game level curiosity, full control, and feeling of independence.

S.5.11.2 The fun factor analysis strategic plan is monitored on a regular basis, and improving it is a continuous strategic effort of the project team.

GDPA 5.12 Ease of Use

S.5.12.1 Consumer feedback indicates satisfaction and ability to navigate conveniently between menu.

S.5.12.2 The defined strategy to enhance consumer experience related to ease of use metrics is regularly reviewed and updated.

GDPA 5.13 Market Orientation

S.5.13.1 The organization  is able to gain competitive advantage by using its market orientation strategy.

S.5.13.2 Developed game concepts are aligned with the requirements of the target market.

S.5.13.3 Developed games are able to maximize their consumers' playing time.

GDPA 5.14 Time to Market

S.5.14.1 Games are published before competitors' games.

S.5.14.2 Being first to market helps to retain existing consumers and attract new ones.

GDPA 5.15 Relationship Management

S.5.15.1 Developed games are able to retain their consumers for a long time.

S.5.15.2 The development team follows a balanced player- and game-centred strategy.

GDPA 5.16 Monetization Strategy

S.5.16.1 The revenue model contributes to strengthening the financial position of the organization.

S.5.16.2 The organization successfully achieves its financial objectives.

S.5.16.3 Return on investment increases over a period of time.

GDPA 5.17 Innovation

S.5.17.1 Past innovative measures taken by the development team have resulted in improved game development and management processes.

GDPA 5.18 Stakeholder Collaboration

S.5.18.1 All stakeholders are involved in game-related decisions.

### 4.4 Performance Scale

The maturity level of an organization is determined by its ability to perform key GDPAs. A five-level scale is used here to rate the maturity level of an organization. A quantitative rating is used to indicate the level of agreement with each statement in the questionnaires and more specifically the way in which the organization fulfils specific maturity-level requirements. Table 3 depicts the ordinal rating used to measure each dimension's GDPAs. The ordinal ratings of a DGMM include "not applicable", "slightly applicable", "partially applicable", "largely applicable" and finally, "completely applicable". Specifically, the rating of "not applicable" is included in a DGMM to increase the flexibility of the methodology. To be consistent with already accepted, validated popular scales such as the BOOTSTRAP methodology [52], the proposed performance scale, and the threshold has been structured accordingly. However, based on the design of a DGMM questionnaires, the linguistic expressions have been slightly adjusted. Overall, the adapted rating methodology and questionnaires are based on the self-assessment approach. This method enables an organization to evaluate its GDPAs by expressing its extent of agreement with the statements.

Table 3: Performance scale

| Scale | Linguistic expression of proposed performance scale | Linguistic expression of BOOTSTRAP | Rating threshold (%) |
|---|---|---|---|
| **4** | Completely applicable | Completely satisfied | ≥ 80 |
| **3** | Largely applicable | Largely satisfied | 66.7-79.9 |
| **2** | Partially applicable | Partially satisfied | 33.3-66.6 |
| **1** | Slightly applicable | Absent/Poor | ≤ 33.2 |
| **0** | Not applicable | - | - |

### 4.5 Rating Method

The rating methodology is adapted from the BOOTSTRAP algorithm [52], as previously mentioned. The rating method consists of a number of different terms such as *development performance rating* ($DPR_{DPA}$), *number of applicable statements* ($NA_{DPA}$), *passing threshold*, ($PT_{DPA}$) and *development maturity level* (DML). Each of these terms is described below in detail.

Let $DPR_{DPA}[a,b]$ be the rating of the $a^{th}$ DPA at the $b^{th}$ maturity level. Subsequently, based on the performance scale described in Table 3, $DPR_{DPA}[a, b]$ can be rated as follows:

**$DPR_{DPA}[a,b]$** = 4 if the extent of applicability of the statement is 80%
= 3 if the extent of applicability of the statement is from 66.7% to 79.9%
= 2 if the extent of applicability of the statement is from 33.3% to 66.6%
= 1 if the extent of applicability of the statement is less than 33.3%
= 0 if the statement is not applicable at all.

The $a^{th}$ statement is considered agreed upon at the $b^{th}$ maturity level:

$$DPR_{DPA}[a,b] \geq 3$$

The number of applicable statements $NA_{DPA}$ at the $b^{th}$ maturity level is defined by the following expression:

$$NA_{DPA}[b] = \text{Number of } \{DPR_{DPA}[a,b] \mid \text{Applicable}\}$$
$$= \text{Number of } \{DPR_{DPA}[a,b] \mid DPR_{DPA}[a,b] \geq 3\}$$

A particular maturity level is considered to be achieved if 80% of the statements in the corresponding questionnaire are applicable to the organization's current status. Table 4 shows the passing threshold for each maturity level, i.e., 80%, rounded to the nearest integer. Hence, if $N_{DPA}[b]$ is the total number of statements at the $b^{th}$ maturity level, then $PT_{DPA}$ at the $b^{th}$ maturity level is defined as follows:

$$PT_{DPA}[b] = N_{DPA}[b] * 80\%.$$

The *game maturity level* (GML) is defined as the highest maturity level at which the number of applicable statements is equal to or greater than $PT_{DPA}$ and is defined as follows:

$$GML = \max\{ b \mid NA_{DPA}[b] \geq PT_{DPA}[b]\}.$$

Table 4: Rating Thresholds.

| Game Maturity Level | Total questions | Passing threshold (80%) |
|---|---|---|
| **Ad-Hoc** | 31 | 25 |
| **Opportunistic** | 51 | 41 |
| **Consistent** | 54 | 43 |
| **Organized** | 54 | 43 |
| **Optimized** | 53 | 42 |

## 5. Proof of Concepts

Case study approach is used as a proof of concept in order to demonstrate the applicability of DGMM in game development organizations. Generally, case study approach is an appropriate strategy, when 'why' or 'how' questions are of primary interest. In this type of research approach, researcher does not have significant knowledge that when the phenomenon of interest will take place or control over different events within some real-life context [58]. The goal of case studies is not to empirically validate a DGMM at this phase of the study, but to demonstrate its applicability in game development industries. The presented DGMM was applied to two game development organizations to assess their game development processes. Details about study design, data collection method and results are described below:

### 5.1 Study Design and Data Collection Strategy

In order to apply developed DGMM in real world, we defined the following research question:

**Research Question:** How can assessment of game development processes be performed within a game development organization by using DGMM?

The assessment questionnaires were designed to assess key process areas and the current maturity level of game development practices. Individuals from participating organizations were requested to select a Likert scale value from 0 to 4 to indicate their extent of agreement with each statement in the questionnaire, as depicted in Table 5. As mentioned in the rating methodology, a statement is considered applicable if the performance rating (according to the method shown in Table 3) is either equal or greater than 3. The respondents to the assessment questionnaires were either project managers or members of the development team from same organizations. All communications with the respondents were carried out either through the survey link or through emails. Participation by respondents was voluntary, and no compensation was offered to them.

In the first stage of the study, contacts were established with many organizations in digital game industry and sent personalized emails to them. After lot of effort, we were able to get positive response from two organizations. The names of the organizations participating in this study are kept confidential due privacy reason. Participating organizations were informed that this assessment formed part of a research study and that subsequently neither the identity of the organization nor of any individual would be disclosed in any publications.

The participating organizations are referred as "Organization A" and "Organization B" in this study. The study participants had been working in the organization for at least three years. In particular, personalized emails were sent stating the objectives of the study. The respondents agreed to participate in the study based on the guarantee that their names or any kind of specific information would not be disclosed in any subsequent publication. Afterwards, they completed the questionnaires for each dimension and level. The questionnaires for each level of maturity serve as a way of obtaining insight into development practices on the identified dimensions of digital game development. The study respondents were requested to provide their extent of agreement with each statement using a performance scale ranging from 0 to 4, as shown in Table 3.

The methodology used for assessment was based on a bottom-up approach, in which respondents had to start from the level 1 questionnaire and then progress to higher levels. The questionnaires for the different levels were also designed using a bottom-up approach, in which more advanced characteristics are introduced as the participant moves from a lower to a higher level.

Organization A is a famous mobile game development company and has released a number of popular games. They are using agile practices for game development. Organization B is another large game development company involved in developing games for various platforms such as mobile, desktop, and Internet. Their game development process is a kind of waterfall with iteration. Table 5 summarizes the assessment results in detail for Organizations "A" and "B".

Both case studies are discussed in detail in the subsequent section. There were multiple responses from each organization, meaning that the chance of bias in the data sample was limited. A number of respondents from each organization (four for organization A and six for organization B), including both project managers and development team members, provided observations about their development practices. In addition, to avoid any chance of biased responses, an inter-rater agreement analysis was performed.

Table 5: Summary of case study assessment results

| Digital Game Maturity Level | Total questions | Passed Threshold 80% | Organization "A" $NA_{DPA}$ | Organization "B" $NA_{DPA}$ |
|---|---|---|---|---|
| **Level 1 (Ad-Hoc)** | 31 | 25 | 29 | 27 |
| **Level 2 (Opportunistic)** | 51 | 41 | 42 | 43 |
| **Level 3 (Consistent)** | 54 | 43 | 44 | 40 |
| **Level 4 (Organized)** | 54 | 43 | 24 | 34 |
| **Level 5 (Optimized)** | 53 | 42 | 18 | 24 |

## 5.2 Organization "A"

Participating organization "A" is one of the leading mobile game development organizations in North America and has developed a number of mobile games. These games include racing, adventure, puzzle, and various role-playing games. Organization "A" can be classified as a medium-sized organization based on its number of employees.

Most of the study participants from Organization "A" had roles that involved policy-making or strategic implementation.

### *5.2.1 Data Analysis*

Once the survey was completed, the $NA_{DPA}$ (number of applicable statements) for each level was calculated. $NA_{DPA}$ was 29 for Level 1, 42 for Level 2, 44 for Level 3, 24 for Level 4, and 18 for Level 5. Therefore, Organization "A" passed the rating threshold of 80% for Level 3, and consequently, Organization "A" is at the "Consistent" level of a DGMM.

In order to get insight into their development practices, we have further analyzed data by using radar charts for each dimension based on their agreement to the statements from 0 to 4 DPR at level 3. Fig. 4 depicts the responses average for each DGMM dimensions at level 3. The chart depicts four dimensions of a DGMM model, which is comprised of DPR ranging from 0 to 4, and it depicts the respondents' agreement level to the statement of each dimension.

Fig. 4 showed clearly that organization 'A' needs to improve specifically its game development methodology dimensions practices, and also have to look into game design strategy dimension and business performance practices.

This will also help them in order to identify their gaps and an idea how they can achieve higher level. In order to dig further into each dimension, for example Fig. 5 depicts the game design strategy spiral chart for all levels of DGMM based on respondents' responses average.

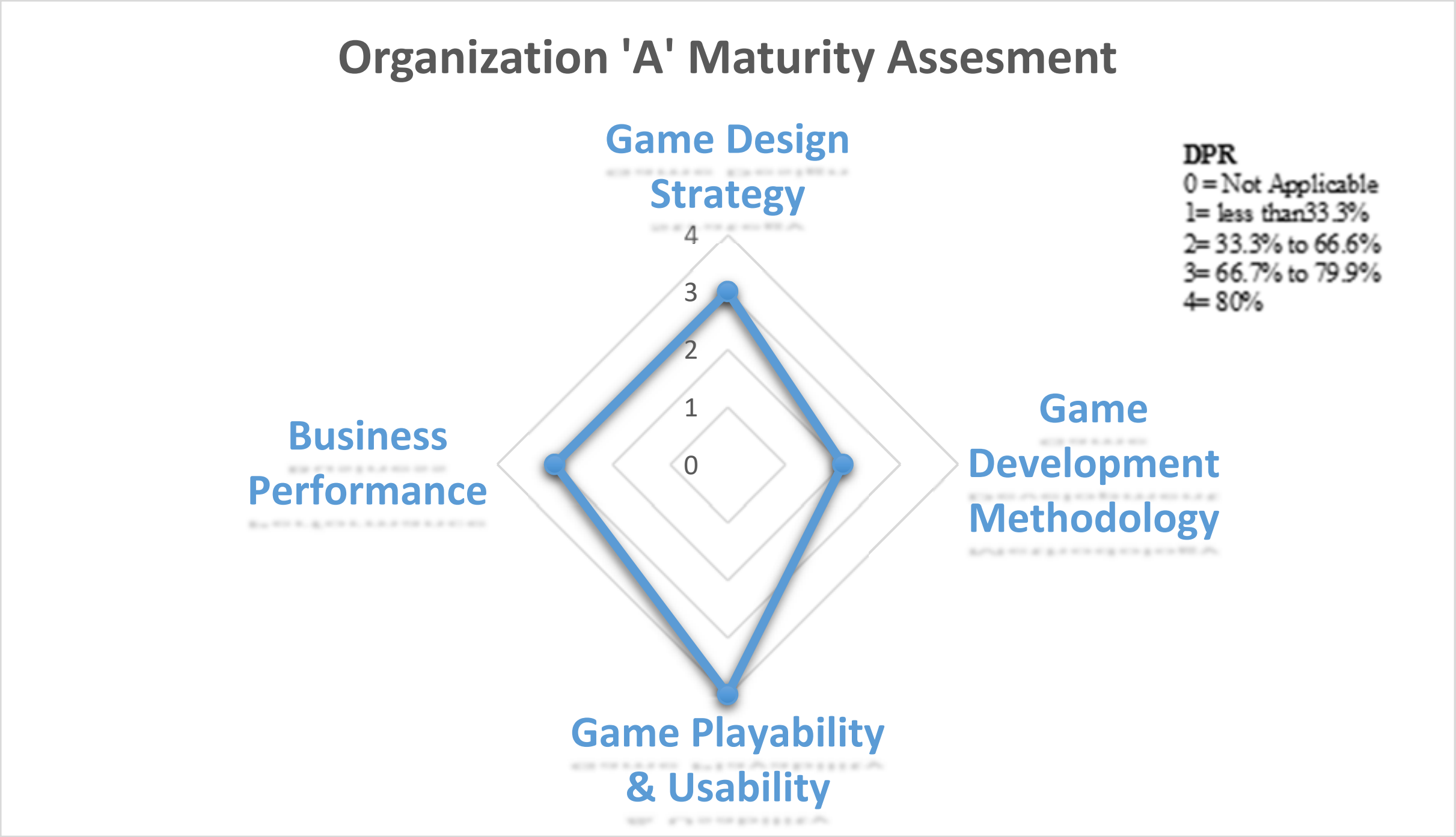

Fig. 4 Organization 'A' DGMM level 3 responses

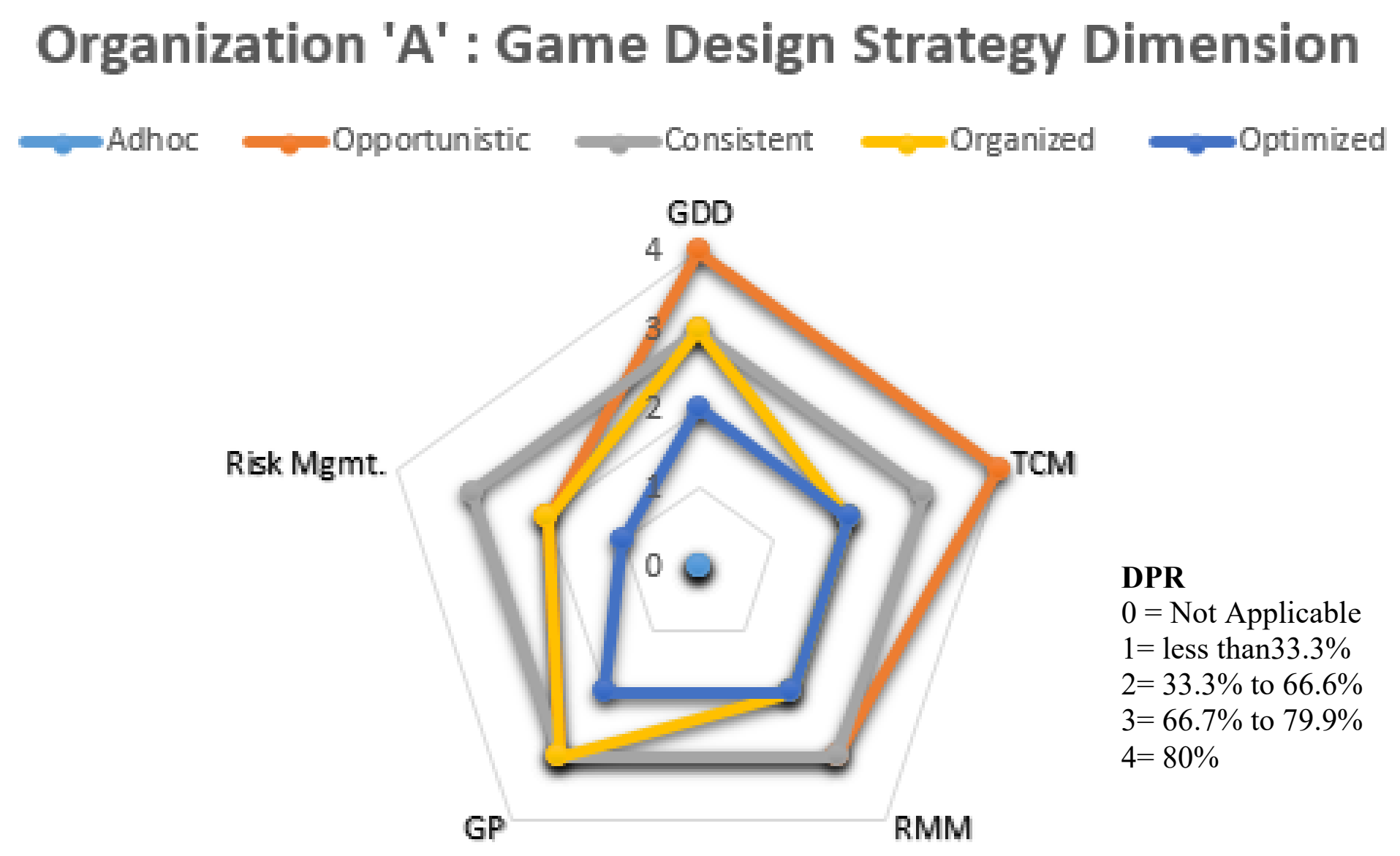

Fig. 5 Responses for Game Design Strategy Dimension process activities for all levels of DGMM of organization A

It clearly shows that under game design strategy dimension, organization 'A' needs to improve its risk management practices to move from level 3 to 4. Moreover, they need to improve their overall game design strategy dimension practices to achieve higher level.

### 5.3 Organization "B"

Organization "B", the second participating organization, is another game development organization in North America that has developed a number of games for various platforms such as mobile, desktop, and Internet. The games can be categorized as role-playing games, adventures, racing games, and puzzles. Organization "B" can be classified as a large organization based on its number of employees. Most of the study participants from Organization "B" had also roles that involved policy-making or strategic implementation.

#### *5.3.1 Data Analysis*

Once the survey was completed, the $NA_{DPA}$ (number of applicable statements) was calculated for each level. $NA_{DPA}$ was 27 for Level 1, 43 for Level 2, 40 for Level 3, 34 for Level 4, and 24 for Level 5. Therefore, Organization "B" passed the rating threshold of 80% for Level 2, and consequently, Organization "B" is at the "Opportunistic" level of a DGMM. In order to analyze data, same strategy is used to get insight into their development practices and identify gaps. Fig. 6 showed the responses averages for each dimension at level 2. The chart depicts four dimensions of a DGMM model, which is comprised of DPR ranging from 0 to 4, and it depicts the respondents' agreement level to the statement of each dimension.

.

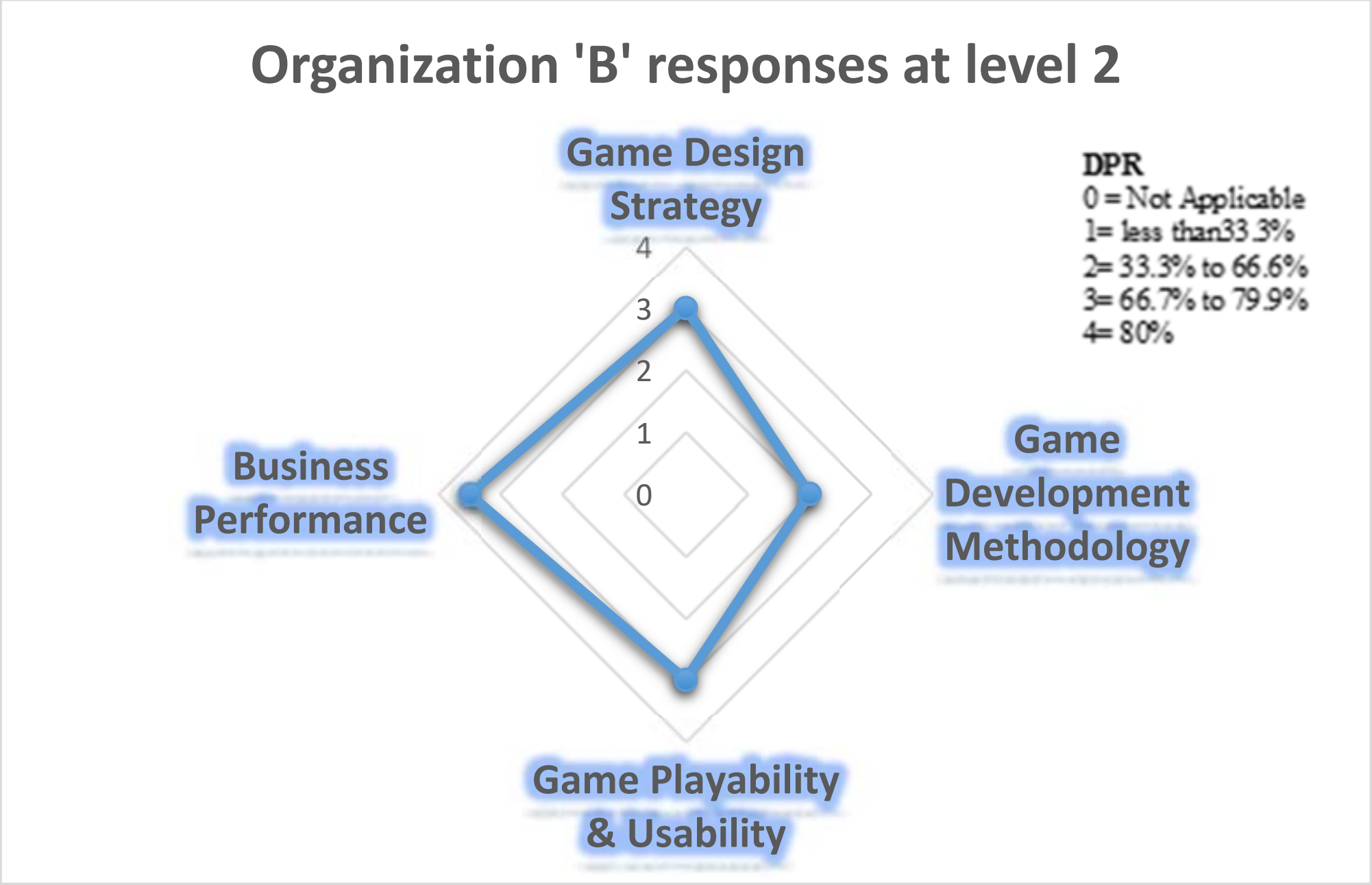

Fig. 6 Organization "B" responses at level 2

It clearly indicates that Organization "B" needs to improve its game development methodology practices to achieve higher levels as mostly respondents' selected DPR 2. Furthermore, they need to improve their development practices for other dimensions. Moreover, Fig 7. depicts the game design strategy dimension results for all levels.

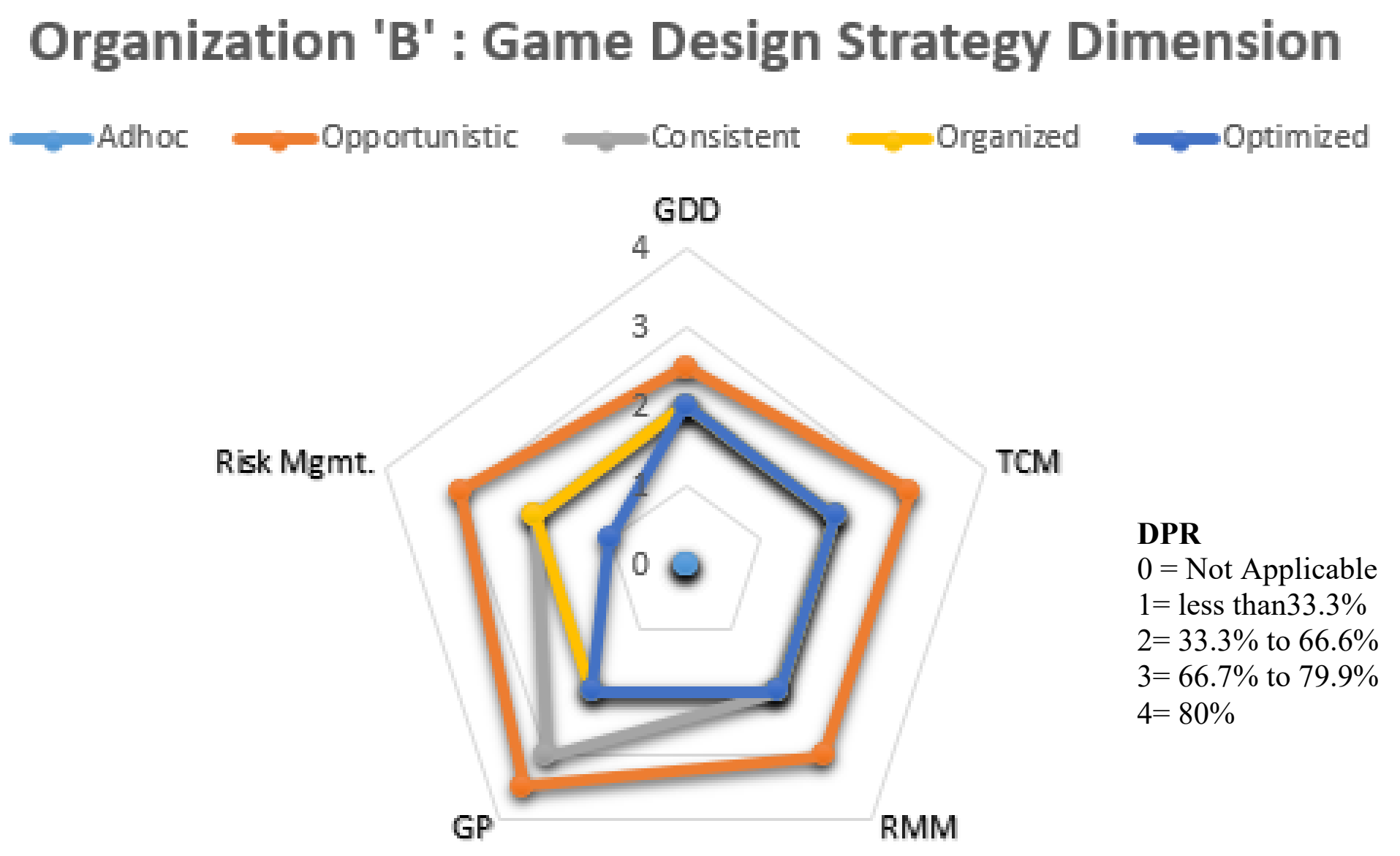

Fig. 7 Responses for Game Design Strategy Dimension process activities for all levels of DGMM of Organization B.

The results clearly depict that "Organization B" needs to improve their GDD management, TCM, RMM and Risk Management practices in order to improve their practices for game design strategy dimensions.

### 5.4 Inter-rater Agreement Analysis

In research such as this, there is generally more than one respondent from each organization, which can create a conflict of opinion about game development practices. To address the issue of conflicting opinion from the same organization, an inter-rater agreement analysis [53] was performed. Inter-rater agreement measures the level of agreement in the ratings provided by different respondents for the same process or software engineering practice [54]. In this case, inter-rater agreement analysis was carried out to identify the level of agreement among different respondents from the same organization. To evaluate inter-rater agreement, the Kendall co-efficient of concordance (*W*) [55] is usually preferred for ordinal data rather than other methods like Cohen's Kappa [56]. "*W*" represents the difference between the actual agreement as drawn from the data and perfect agreement. Values of Fleiss Kappa and the Kendall's *W* coefficient can range from 0 (representing complete disagreement) to 1 (representing perfect agreement) [57]. Therefore, the Kappa [54] standard includes four levels: < 0.44 means poor agreement, 0.44 to 0.62 represents moderate agreement, 0.62 to 0.78 indicates substantial agreement, and > 0.78 represents excellent agreement. In this study, the observed Kappa coefficient fell into the substantial category, ranging from 0.63 to 0.68. Table 6 reports the Kappa and Kendall statistics for both organizations; both measures fall into the "substantial" category.

Table 6: Inter-rater agreement analysis of Organizations "A" and "B".

| DGMM level | Organization A | | | | Organization B | | | |
|---|---|---|---|---|---|---|---|---|
| | *Kendall's Coefficient of Concordance* | | *Fleiss statistics* | *Kappa* | *Kendall's Coefficient of Concordance* | | *Fleiss statistics* | *Kappa* |
| | *Coef.(W)* | $X^2$ | *Coef.* | *Z* | *Coef.(W)* | $X^2$ | *Coef.* | *Z* |
| **Ad-Hoc** | 0.65 | 52.91* | 0.69 | 8.91* | 0.85 | 53.44* | 0.75 | 4.49* |
| **Opportunistic** | 0.72 | 58.10* | 0.67 | 7.99* | 0.71 | 55.31* | 0.67 | 5.03* |
| **Consistent** | 0.63 | 52.13* | 0.66 | 8.20* | 0.89 | 55.62* | 0.64 | 4.88* |
| **Organized** | 0.71 | 56.74* | 0.62 | 7.66* | 0.98 | 47.66* | 0.78 | 5.29* |
| **Optimized** | 0.73 | 57.20** | 0.63 | 7.01** | 0.88 | 53.48** | 0.80 | 5.01* |

Significant at $p<0.01$*; Significant at $p<0.05$**.

## 6. Discussion & Limitations of study

In software engineering, maturity models make it possible to obtain comprehensive insight into current development processes, their related activities, and their current level of maturity. This information can be useful in streamlining current strategic plans and improving future activities. Furthermore, maturity models enable an organization to position itself and provide motivation to adopt good practices and strive for the next level.

Game development proved to be an incredibly challenging topic of research because game technology, including game platforms and engines, changes rapidly, and coded modules are very rarely used in another game project. However, the recent success of the digital game industry has imposed further stress along with game development challenges and highlights the need to adopt good game development practices. This will enable organizations to meet player demands and be successful in the highly competitive digital game industry. Ultimately, it will make it easier for organizations to meet their financial objectives. To determine the maturity of a current process or a specific area in the game development process that needs improvement, an assessment of process activities must be performed. However, due to the relatively short history and empirical nature of the field, the topic of development strategies and best practices for game development has not been fully explored.

To identify the important dimensions of digital game development methodology, empirical investigations and literature reviews were carried out to examine the impact of key factors on game development [17]. Based on an examination of key factors, research models have been developed to perform studies. Relationships were then established between key game development factors and the perspectives of different key stakeholders. To assess game development practices, the significant key factors identified from three empirical investigations have been used as a measuring instrument to develop a DGMM. The structural composition of DGMM comprised of a four-dimensional assessment framework based on the perspectives of game developers, consumers, and businesses.

Subsequently, a DGMM as developed has been used to assess the level of maturity of current development practices by following the developed assessment methodology and conducting case studies. Specifically, the model presented here can be used by an organization to assess its current practices and enable it to discover bottlenecks in its current methodology. Finally, it can be used to improve current processes and provide opportunities to develop successful games. The overall performance of a digital game is dependent on how it handles the usability and playability factors to attract its players, as well as how well developers meet deadlines by following suitable game design and development strategies. Finally, game performance has an impact on the business dimension of an organization. The DGMM presented here will help organizations to streamline their current practices, which can lead to the completion of more successful game development projects This will also helps the development and management team to obtain insight into current development practices and to identify specific key process areas to improve. Game development processes are complex and need careful monitoring and evaluation to meet the organization's objectives. The proposed DGMM provides an early conceptual framework for maturity assessment of game development practices. Further contributions in this particular area are still required from academic and industrial researchers.

The assessment methodology of the proposed DGMM is based on questionnaires and hence is susceptible to certain limitations. Although the proposed maturity model is based on three empirical studies involving five maturity levels and eighteen different key factors, it is always possible that other factors such as game category, organization size, and cultural and economic conditions have been inadvertently excluded. The proposed model assessment methodology is based on subjective responses from project managers and development team members. However, approaches used to ensure reliability and validity are part of the common statistical techniques used by software engineering researchers. Another limitation of the model is that there is a chance of biased response from respondents because measuring instrument is presented using bottom up approach. For example, if a respondent say that GDD is seen helpful in preproduction phase but also claims that they don't develop GDD, similarly, a scenario when the development team want to hide their poor development practices from project manager. The possible solution to this limitation is to remove such responses from data to ensure the validation of the instrument or to present the measuring instrument using randomized approach.

Generally, independent assessors are considered essential in defining the coordination with the internal assessment team. However, the proposed methodology does not consider the role of an independent assessor, and case studies are performed based on self-assessment. The proposed DGMM provides numerical data regarding the level of maturity of game development practices and factors within organizations, but it does not provide any guidelines to improve them. Guidelines for process improvement and how to progress from lower to higher levels remain topics for further research.

Although the model as presented here has certain specific and general limitations, key game development factors have been validated using commonly used statistical approaches. This study provides a comprehensive approach to evaluate current game development practices and give organizations insight into their development activities. This study also provides future directions for research in game development.

## 7. Final remarks

Game development has interesting properties such as real-time interaction, emergence, and computationally challenging components that create a new field of study despite its similarity to software engineering. Software engineering practices are also beneficial for the game development industry, but the industry largely still discounts software engineering practices as ultimately unable to meet its needs when working on complex development tasks. To aid in facing the challenges encountered by game development organizations, solutions will have to satisfy developers as well as consumers and must be applicable to general game development. They must not be specific to one game genre or cultural environment. Hence, solutions must incorporate best software engineering practices plus project management skills so that game investors and game players do not need to worry about the quality of their game software product because of its reusability, reliability, or expandability.

Assessment of game development processes is an important area of research for developing quality games and ultimately for their successful business performance. Currently, no research has been reported in the area of game development methodology assessment. This paper has proposed a DGMM that includes key game development factors and crucial concepts from software engineering and project management. It is easily applicable to any size of the organization, any game genre, and any platform. A DGMM provides a set of best practices for managing complex game projects. The proposed model can be effectively used and tailored for the development of any kind of game project. A DGMM framework consists of assessment questionnaires for the five maturity levels, a rating method, and performance scales. In the course of this research, case studies were conducted to demonstrate the methodology for evaluating the level of maturity of game development in two organizations.